\documentclass{ws-ijmpe}

\usepackage{url}
\usepackage[super,compress]{cite}

\usepackage{hyperref}
\usepackage{breakurl}
\usepackage{color}
\usepackage{outlines}

\begin{document}

\markboth{A. Dohi}{Cooling of Isolated Neutron Stars with Pion Condensation}

\catchline{}{}{}{}{}

\title{Cooling of Isolated Neutron Stars with Pion Condensation:\\
Possible Fast Cooling in a Low-Symmetry-Energy Model}

\author{Akira Dohi}
\address{Department of Physics, Kyushu University, Fukuoka 819-0395, Japan\\
Interdisciplinary Theoretical and Mathematical Sciences Program (iTHEMS), RIKEN, Wako 351-0198, Japan\\
dohi@phys.kyushu-u.ac.jp }
\author{Helei Liu}
\address{School of Physical Science and Technology, Xinjiang University, Urumqi 830046, China}
\author{Tsuneo Noda}
\address{Kurume Institute of Technology, Fukuoka 830-0052, Japan}
\author{Masa-aki Hashimoto}
\address{Department of Physics, Kyushu University, Fukuoka 819-0395, Japan}

\maketitle

\begin{history}
\end{history}

\begin{abstract}%
We studied thermal evolution of isolated neutron stars (NSs) including the pion condensation core, with an emphasis on the stiffness of equation of state (EOS). Many temperature observations can be explained by the minimal cooling scenario which excludes the fast neutrino cooling process. However, several NSs are cold enough to require it. The most crucial problem for NS cooling theory is whether the nucleon direct Urca (DU) process is open. The DU process is forbidden if the nucleon symmetry energy is significantly low. Hence, another fast cooling process is required in such an EOS. As the candidate to solve this problem, we consider the pion condensation. We show that the low-symmetry-energy model can account for most cooling observations including cold NSs, with strong neutron superfluidity. Simultaneously, it holds the $2~M_{\odot}$ observations even if the pion condensation core exists. Thus, we propose the possibility of pion condensation, as an exotic state to solve the problem in low-symmetry-energy EOSs. We examined the consistency of our EOSs with other various observations as well.
\end{abstract}


\ccode{PACS numbers: 26.60.--c, 26.60.Kp}


\newcommand{\apj}{Astrophys. J. }
\newcommand{\apjs}{Astrophys. J. Suppl. }
\newcommand{\apjl}{Astrophys. J. Lett. }
\newcommand{\pasj}{Publ. Astron. Soc. Japan. }
\newcommand{\pasa}{Publ. Astron. Soc. Australia. }
\newcommand{\physrep}{Phys. Rep. }
\newcommand{\ptp}{Prog. Theor. Phys. }
\newcommand{\ptps}{Prog. Theor. Phys. Suppl. }
\newcommand{\ptep}{Prog. Theor. Exp. Phys. }
\newcommand{\AIP}{AIP Conf. Proc. }
\newcommand{\aap}{Astron. Astrophys. }
\newcommand{\ssr}{Space Sci. Rev. }
\newcommand{\sci}{Science }
\newcommand{\prc}{Phys. Rev. C }
\newcommand{\prd}{Phys. Rev. D }
\newcommand{\prl}{Phys.~Rev.~Lett.~}
\newcommand{\nar}{New Astron. Rev. }

\newcommand{\araa}{Ann. Rev. Astron. Astrophy. }
\newcommand{\mnras}{Mon. Not. Roy. Astron. Soc. }

\newcommand{\nphysa}{Nucl. Phys. A }

\newcommand{\jcap}{JCAP}

\newcommand{\memsai}{Memorie della Soc. Astron. Ital. }

\section{Introduction}
\label{intro}

Neutron stars (NSs) are born just after a supernova explosion of a massive progenitor $<20 M_{\odot}$. After that, it is believed that the hot NS cools down by the losses of neutrinos and photons. For $t\lesssim10^{5}~{\rm yrs}$ after the formation of NSs, the cooling behavior is dominated by neutrinos. Since the cooling behavior is naturally reflected on the temperature of NSs, the observed surface temperature may provide information on the equation of state (EOS), which is still uncertain in ultrahigh density regions. Recent observations of NS mass and radius, such as high-mass millisecond pulsars~\cite{2010Natur.467.1081D,2013Sci...340..448A,2020NatAs...4...72C} and GW170817~\cite{2017PhRvL.119p1101A,2018PhRvL.121p1101A}, impose some constraints on the EOS (in particular $P$--$\rho$ relation). Therefore, cooling observations of NSs could give even stronger constraints including the composition.

The influence of EOS uncertainties on cooling curves has been discussed for many years. In particular, whether NSs cool rapidly is an important question, which is determined by the EOS. Such a fast $\nu$ cooling process is forbidden in relatively light NSs without any exotic state. In such normal stars, cooling curves are dominated by slow cooling processes (modified Urca and the bremsstrahlung) and enhanced cooling process which is caused by the nucleon superfluid state (pair breaking formation; PBF). Such a so-called {\it minimal cooling scenario} is believed to explain the most of isolated NSs observations~\cite{2004ApJS..155..623P,2004A&A...423.1063G,2009ApJ...707.1131P}.

However, some observed NSs are too cold to be clearly explained by minimal cooling scenario. For example, focusing on isolated NSs, PSR J0205+6449 in supernova remnant 3C58 and RX J0007.0+7302 are known to be too cold for their young ages~\cite{2004ApJS..155..623P}. Other compact objects such as G127.1+0.5, G084.2$+$0.8, G074.0$-$8.5, G065.3$+$5.7~\cite{2004ApJS..153..269K}, and G043.3$-$0.2~\cite{2006ApJS..163..344K} are extraordinary cold from X-ray observations~\cite{2004ApJS..153..269K,2006ApJS..163..344K}, although they have not been identified yet~(see also Figure 11 in Ref.~\cite{2020ApJ...888...97B}). When we focus on the cold accreting NS observations, SAX J1808.4$-$3658 and 1H 1905+000 are very faint in a quiescent period, despite their high accretion rate~\cite{2009ApJ...691.1035H}. In addition, recent observation of outbursts in the transient system MXB 1659$-$29 suggests that the accreting NS is extraordinary cold~\cite{2018PhRvL.120r2701B}. To reproduce such observations of cold stars, fast cooling processes are necessary. Such a fast cooling process does not occur unless the momentum conversation law of the particles involved is satisfied. For the nucleon direct Urca (DU) process as an example, the condition is 
determined by proton fraction $Y_p$ and the threshold is 1/9 (without muons)~\cite{1981PhLB..106..255B,1991PhRvL..66.2701L}. If $Y_p$ exceeds 1/9 and the mass exceeds the corresponding mass $M_{\rm DU}$ , the fast cooling process occurs and cools the NSs rapidly. Thus, EOS is an important factor to determine the mass which distinguishes the NS cooling scenario.

The stiffness of the EOS and the behavior of cooling curves are connected with the nucleon symmetry energy, which corresponds to the proton fraction $Y_p$. The density dependence of the symmetry energy is characterised by several parameters as a representative of the slope parameter $L$~\cite{2009PhRvC..80d5806V}. From some previous studies, it is shown that $L$ is correlated with both the NS radius and  $M_{\rm DU}$~\cite{2017IJMPE..2650015L,2019PTEP.2019k3E01D}. Since $L$ has been currently constrained to be below $80~{\rm MeV}$~\cite{2014EPJA...50...40L}, the possibility of the DU process in light stars can be excluded. In other words, heavy stars are likely to cool rapidly (but see also Ref.~\cite{2013ApJ...765....1N}). However, if the symmetry energy is significantly low, the DU process itself is forbidden with any masses.  As one of such EOSs, we focus on the TOGASHI ($L=30.0~{\rm MeV}$), which has been recently constructed based on realistic two-body interaction and phenomenological three-body interaction~\cite{2017NuPhA.961...78T}. In this EOS, the radius with $1.4 M_{\odot}$ stars is 11.6 km, which matches with some recent observations~(e.g., Refs.~\cite{2010ApJ...722...33S,2018PhRvL.121p1101A}). The $L$ value of the TOGASHI is relatively low but is still consistent with the experimental and observational constraints of according with Ref.~\cite{2017RvMP...89a5007O} (but see also Ref.~\cite{2020PhRvL.125t2702D} for more latest constraints). However, the TOGASHI cannot explain the cold NSs~\cite{2019PTEP.2019k3E01D}. In such a low-symmetry-energy EOS, another fast cooling process involving exotic matter is required. There are several candidates of exotic states such as hyperon mixing, quark deconfinement, and meson($\pi^-,K^-$) condensation. Here we consider the pion condensation as one of the possible solutions. 

The possibility of pion condensation in the dense matter and NS core has been discussed for decades (for review, see Refs.~\cite{1978RvMP...50..107M,2020PAN....83..188V}). In this idea, the pion condensation is induced by the coherent ground states with the same quantum number, spin, and isospin. The pion condensation occurs over nuclear saturation density $\rho_{\rm nuc}$($=2.66\times10^{14}~{\rm g~cm^{-3}}$ for the TOGASHI), though the appearance density are still unclear. If the pion condensation occurs, the momentum of quasi-particles becomes large enough to cause the DU process. Then, the strong pion Urca process can occur and significantly reduce the surface temperature. Hence, the pion condensation is a candidate to explain cold NS stars~\cite{1994ApJ...431..309U,2018IJMPE..2750067M}. A remarkable feature of pion condensation is that the appearance density might be low with $\rho_{\rm B}\approx2\rho_{\rm nuc}$ unlike other exotic states~\cite{2005PhLB..615..193Y,2006PrPNP..56..446I}. This implies that a fast cooling process is assured to occur, although the EOS becomes greatly soft. That is why the original EOS without the pion condensation effect must become stiff enough to easily satisfy recent observations. Focusing on the stiffness of standard-matter EOS, we examine the structure and cooling of isolated NSs with pion condensation.

This paper is structured as follows: In Sect.~\ref{sec:eos}, we present the EOS and mass-radius relation with pion condensation, including the test of our EOSs with several observations. In Sect.~\ref{sec:nu}, we briefly explain our formulation to calculate cooling curves. In Sect.~\ref{sec:result}, we present the cooling curves with some constructed EOSs with pion condensation, and discuss the consistency with cooling observations. In Sect.~\ref{sec:conc}, we finally give a conclusion.  

\section{Equation of State with Pion Condensation}
\label{sec:eos}

As the EOSs without exotic states, we adopt the TOGASHI~\cite{2017NuPhA.961...78T}, focused in this study. As the comparison, we also adopt the other EOSs of TM1~\cite{1998NuPhA.637..435S,1998PThPh.100.1013S,2011ApJS..197...20S} and TM1e~\cite{2019ApJ...887..110S,2020ApJ...891..148S}. They have been constructed based on the relativistic mean-field theory with several meson coupling terms. The only difference is that the TM1e includes $\omega$-$\rho$ coupling term, unlike the TM1. By considering such a term, the EOS is known to become softer around $\rho_{\rm nuc}$~\cite{2020ApJ...891..148S}. The TM1 and TM1e have $L=111.1~{\rm MeV}$ and $40~{\rm MeV}$, respectively. Compared with these EOSs, the symmetry energy with the TOGASHI is very low enough to prohibit the DU process with any masses. In a low-density region around $\rho<10^{11}~{\rm g/cm^3}$, we connect the adopted EOS to the BPS EOS~\cite{1971ApJ...170..299B} for the outer layer.

As the nuclear model of pion condensation, we adopt Ref.~\cite{1993PThPS.112..221M} based on SU(2) chiral symmetry approach. This model incorporates realistic interactions such as the attractive force by isobar $\Delta(1232)$ excitations and the repulsive force by baryon-baryon short-range correlations. Since their interactions are in conflict with each other, the competitive relationship of nuclear force arises. Then, if the total nuclear interaction becomes attractive, the pion condensation occurs. The appearance density is characterised by Landau-Migdal parameter $\tilde{g}^{\prime}$,
which means the strength of nucleon-nucleon($\tilde{g}^{\prime}_{NN}$), nucleon-isobar($\tilde{g}^{\prime}_{N\Delta}$), and isobar-isobar($\tilde{g}^{\prime}_{\Delta\Delta}$) interactions, assuming universality as $\tilde{g}^{\prime}\equiv \tilde{g}^{\prime}_{NN}=\tilde{g}^{\prime}_{N\Delta}=\tilde{g}^{\prime}_{\Delta\Delta}$.
We adopt $\tilde{g}^{\prime}=0.5$ in our models. Charged pion ($\pi^c$) condensation phase appears at $\rho_{\rm B}\simeq1.6\rho_{\rm nuc}$. In much higher density regions, $\pi^c$ condensation phase transits into a combined phase of neutral and charged pions ($\pi^0$--$\pi^c$) condensation at $\rho_{\rm B}\simeq3.9\rho_{\rm nuc}$~\cite{1987PThPh..78.1405M} (see Table~5 in Ref.~\cite{1994ApJ...431..309U}). Although the short-range correlation in nuclei is highly uncertain, some experiments have indicated that the pion condensation begins at $(1.9\pm0.3)\rho_{\rm nuc}$ or (1.8$-$2.4)$\rho_{\rm nuc}$~\cite{2005PhLB..615..193Y,2006PrPNP..56..446I}. These experimental results agree with the theoretical prediction of Ref.~\cite{1993PThPS.112..221M}. If the pion condensation occurs, the EOS becomes significantly soft. The quantitative effect depends on the stiffness of properties with the standard-nuclear matter without pion condensation phases. Focusing on this point, we investigate how the stiffness of standard-nuclear EOS changes the pressure-density and mass-radius relations with pion condensation. 

\begin{figure}[t]
    \centering
    \includegraphics[width=0.8\linewidth]{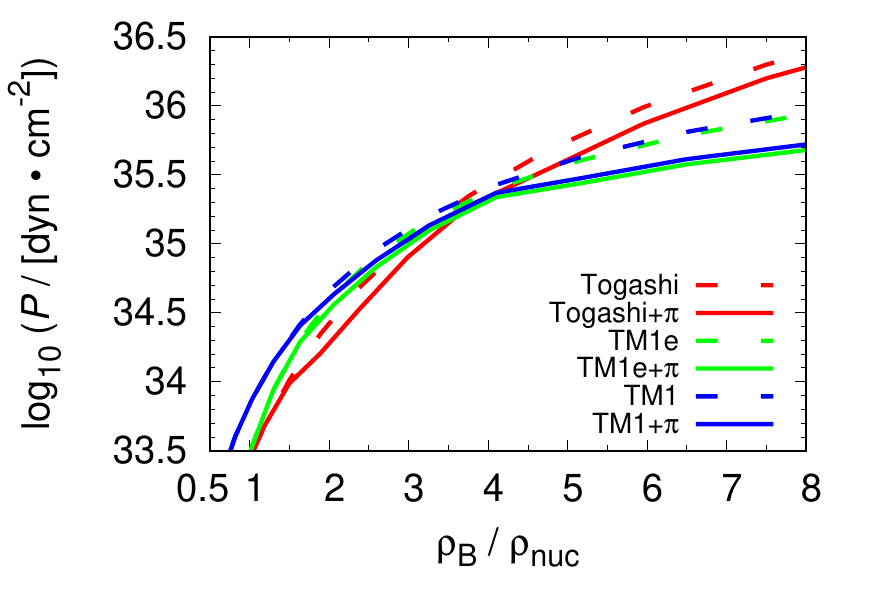}
    \caption{Pressure-Density relation with adopted EOSs: TOGASHI(red), TM1e(green), and TM1(blue). Solid curves indicate EOS without pion condensation while Dashed curves with pion condensation modeled by Ref.~\cite{1993PThPS.112..221M} (labeled as ``EOS$+\pi$").}
    \label{fig:prho}
\end{figure}
 
By adding the original energy density and pressure on their gains due to the pion condensation, we construct the EOS with pion condensation. The values of the gains are listed in Table~1--4 of Ref.~\cite{1994ApJ...431..309U}.
At first, we show the pressure-density relations of the constructed EOSs in Fig.~\ref{fig:prho}. As we see, the TOGASHI is soft for $\rho\sim\rho_{\rm nuc}$ and stiff for $\rho\gg\rho_{\rm nuc}$. This trend is opposite from the cases of TM1 and TM1e. Considering the pion condensation, softening effect appears at lower density with the TOGASHI compared with others. Meanwhile, the TOGASHI does not become softer for $\rho_{\rm B}\gtrsim4\rho_{\rm nuc}$ compared with other EOSs. 

\begin{figure}[t]
\centering
    \includegraphics[width=0.8\linewidth]{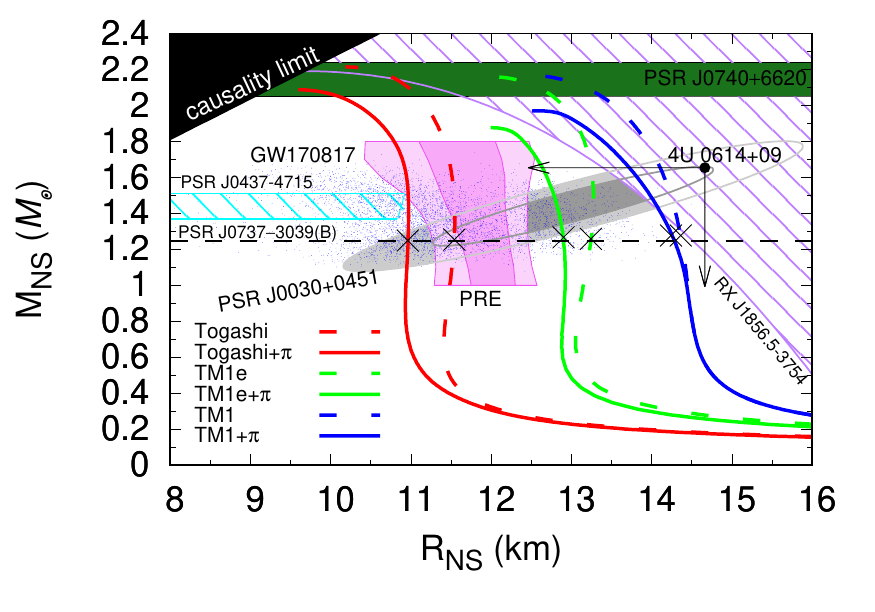}
    \caption {Mass-radius relation with adopted EOSs. The green band shows the highest-mass measurement of $2.14^{+0.10}_{-0.09} M_{\odot}$ of PSR J0740--6620~\cite{2020NatAs...4...72C}. Many blue dots indicate the results from the observation of GW170817~\cite{2018PhRvL.121p1101A}. Grey cycles indicate the limits of allowed $1\sigma$(thick) and $2\sigma$(thin) regions of pulsar J0030+0451 discovered by {\it NICER} observation ~\cite{2019ApJ...887L..24M,2020Univ....6...81B}. Thick($1\sigma$) and thin($2\sigma$) magenta regions show the allowed ones from observational constraints of photospheric radius expansion in six low-mass X-ray binaries (LMXBs)~\cite{2010ApJ...722...33S}. The black point shows the upper-limit mass and radius implied from the kHz QPO observations in LMXBs 4U 0614+09~\cite{2000ApJ...540.1049V}. Two hatched regions labeled by purple and cyan indicate the excluded ones derived from the thermal radiation of the isolated NS RX J1856.5–3754~\cite{2002ApJ...564..981P,2004NuPhS.132..560T} and thermal pulses from a radio millisecond pulsar PSR J0437--4715~(3$\sigma$ regions in Ref.~\cite{2013ApJ...762...96B}), respectively. For the latter constraint, we combine it with the observed mass of $M_{\rm NS}=1.44\pm0.07~M_{\odot}$ reported by Ref.~\cite{2016MNRAS.455.1751R}. ``$\times$" denotes the mass with the baryon mass $M_{\rm B}=1.375~M_{\odot}$ which is the upper limit from observation of PSR J0737$-$3039(B). Its lower limit of observational mass $M_{\rm NS}=1.248~M_{\odot}$ is shown as the dashed black line. If the observational mass is higher than the mass denoted as ``$\times$", the EOS is rejected.}
    \label{fig:mr}
\end{figure}

Based on the constructed EOSs, we consider the mass-radius relation, which is obtained by solving Tolman-Oppenheimer-Volkoff(TOV) equations~\cite{1939PhRv...55..364T,1939PhRv...55..374O}. The results are present in Fig.~\ref{fig:mr}. We also show the maximum mass and radius with some masses in Table~\ref{tab:eosmr}. The pion condensation highly decreases the maximum mass and radius, and their density dependence is different among EOSs. Stiff EOSs for $\rho_{\rm B}\gg\rho_{\rm nuc}$ do not feel the softening effect due to pion condensation so much as we see Fig.~\ref{fig:prho}. Hence, the maximum mass with the TOGASHI is not significantly reduced due to the pion condensation, while highly reduced with other EOSs. On the other hand, the TOGASHI is soft around $\rho_{\rm B}\sim \rho_{\rm nuc}$ and the radius is highly reduced even with light stars. These trends are opposite with the stiff EOSs such as the TM1. The TM1e has roughly intermediate
property between TOGASHI and TM1 ones.

\begin{table}[t]
    \centering
    \caption{Maximum mass $M_{\rm max}$ and radius with different masses for each EOS.}
    \scalebox{0.85}{
    \begin{tabular}{c|ccccc}
    \hline\hline
    EOS & $M_{\rm max}$ ($M_{\odot}$) &
    $R_{\rm NS}(1.0 M_{\odot})$ (km) &
    $R_{\rm NS}(1.4 M_{\odot})$ (km)  & $R_{\rm NS}(1.8 M_{\odot})$ (km)  & $R_{\rm NS}(M_{\rm max})$ (km)\\
    \hline
      TOGASHI   & 2.21 & 11.49 &11.57 & 11.41 &10.18\\
    TOGASHI+$\pi$ & 2.09 & 10.94 & 10.97 & 10.79 & 9.60  \\
      TM1e   & 2.13 & 13.15& 13.27 & 13.16 & 12.40\\
      TM1e+$\pi$   & 1.88 & 12.92 & 12.85 & 12.40 & 11.99\\
       TM1   & 2.18 & 14.45 &14.30 & 13.93 & 12.35\\
      TM1+$\pi$   & 1.97 & 14.44 & 14.12 & 13.40 & 12.51\\
    \end{tabular}
    }
    \label{tab:eosmr}
\end{table}

Next, we compare our EOSs with various observational constraints. The adopted constraints are briefly explained in subsections \ref{subsec:a}--\ref{subsec:i}. The final results for the evaluation of our EOSs are summarised in Table~\ref{tab:summary}.

\begin{table*}[t]
    \centering\hspace*{-1.0cm}
      \caption{Summary of the results to test our EOSs with use of observational constraints present in subsections \ref{subsec:a}--\ref{subsec:i}. }
    \scalebox{0.9}{
    \begin{tabular}{|c|c|c|c|c|c|c|c|c|c|c|c|c|}
    \hline \
       &  \ref{subsec:a}
        & \ref{subsec:b}  & \multicolumn{2}{c|}{\ref{subsec:c}} & \multicolumn{2}{c|}{\ref{subsec:d}} &
        \multicolumn{1}{c|}{\ref{subsec:e}} &
    \ref{subsec:f} & \ref{subsec:g} & \ref{subsec:h} & 
    \ref{subsec:i} \\
      EOS
      & ($M_{\rm max}\geqslant2.05 M_{\odot}$)
      &
      & $1\sigma$ & $2\sigma$ 
       & $1\sigma$ & $2\sigma$
      & ($3\sigma$)
      &
      & 
      &  
      & ($\Delta I/I\ge1.8\%$)\\ \hline
      TOGASHI
      & $\checkmark$
      & $\checkmark$
      & $\checkmark$
      & $\checkmark$
      & $\checkmark$
      & $\checkmark$
      & $\checkmark$
      & $\checkmark$
      & {\small $M_{\rm NS}\lesssim2.1 M_{\odot}$}
      & $\checkmark$
      & {\small $M_{\rm NS}\le1.82 M_{\odot}$} \\
       TOGASHI+$\pi$
      & $\checkmark$
      & $\checkmark$
      & $\times$
      & $\checkmark$
      & $\times$
      & $\checkmark$
      & $\checkmark$
      & $\checkmark$
      & $\checkmark$
      & $\checkmark$
      & {\small $M_{\rm NS}\le1.69 M_{\odot}$} \\
      TM1e
      & $\checkmark$
      & $\checkmark$
      & $\checkmark$
      & $\checkmark$
      & $\times$
      & $\times$
      & $\checkmark$
      & $\checkmark$
      & {\small $M_{\rm NS}\lesssim1.7 M_{\odot}$}
      & $\checkmark$
      & $\checkmark$ \\
      TM1e+$\pi$
      & $\times$
      & $\checkmark$
      & $\checkmark$
      & $\checkmark$
      & $\times$
      & $\checkmark$
      & $\checkmark$
      & $\checkmark$
      & $\checkmark$
      & $\checkmark$
      & {\small $M_{\rm NS}\le1.87 M_{\odot}$} \\
      TM1
      & $\checkmark$
      & $\times$
      & $\checkmark$
      & $\checkmark$
      & $\times$
      & $\times$
      & $\checkmark$
      & $\checkmark$
      & {\small $M_{\rm NS}\lesssim1.3 M_{\odot}$}
      & $\checkmark$ 
      & $\checkmark$ \\
      TM1+$\pi$
      & $\times$
      & $\times$
      & $\checkmark$
      & $\checkmark$
      & $\times$
      & $\times$
      & $\checkmark$
      & $\checkmark$
      & {\small $M_{\rm NS}\lesssim1.4 M_{\odot}$}
      & $\checkmark$ 
      & $\checkmark$ \\ \hline
    \end{tabular}
    }
    \label{tab:summary}
\end{table*}

\subsection{Maximum mass constraint from a heaviest object of PSR J0740$+$6620}
\label{subsec:a}
Recent observations of some massive pulsars exclude many soft EOSs which cannot support $2 M_{\odot}$. In particular, if exotic states are included, the EOS tends to significantly become soft, and may fail to reproduce $2 M_{\odot}$ stars ({\it hyperon puzzle}, but see also Ref.~\cite{2021arXiv210606687K}). Hence, {\it hyperon puzzle} is the most crucial condition for the test of exotic-matter EOSs. Currently, the heaviest NS observed so far is a pulsar PSR J0740+6620. The Shapiro-delay based mass measurement shows $M=2.14^{+0.10}_{-0.09}M_{\odot}$~\cite{2020NatAs...4...72C}.
In adopted EOSs, the TM1e$\pi$ and TM1$+\pi$ are inconsistent with the observation, while the TOGASHI$+\pi$ is consistent. As the above, the main reason is the difference in stiffness of standard-nuclear-matter EOS in high-density regions. Hence, considering the low-symmetry-energy EOS with pion condensation has a high possibility to reproduce $2M_{\odot}$ stars.

\subsection{Constraint from GW170817}
\label{subsec:b}
The first discovery of gravitational wave emitted from NS-NS merger, GW170817, gives the constraint on the radius. The key to extracting the information of interior NS from the analysis of gravitational-wave data is how the NSs are deformed by the tidal force, which is estimated to be $\Lambda_{1.4}\lesssim800$~\cite{2017PhRvL.119p1101A} or $\Lambda_{1.4}\lesssim580$~\cite{2018PhRvL.121p1101A} where $\Lambda_{1.4}$ is the tidal deformability with $1.4~M_{\odot}$ stars. The data of mass-radius constraints are adopted from Ref.~\cite{2018PhRvL.121p1101A}, which results in $R_{\rm NS}\lesssim13.6~{\rm km}$ with $1.4 M_{\odot}$ stars. Hence, many EOSs with high symmetry energy, such as the TM1, are rejected. 

\subsection{Constraint from recent {\it NICER} observation of PSR J0030$+$0451}
\label{subsec:c}
Recently, the X-ray timing observation of a millisecond pulsar PSR J0030+0451 by Neutron Star Interior Composition Explorer ({\it NICER}) enables us to constrain the EOS, in particular, the compactness $M/R$ ~\cite{2019ApJ...887L..24M}. The detailed analysis considering the correlation between mass and radius has been done by Ref.~\cite{2020Univ....6...81B}, which results in $M_{\rm NS}=1.44\pm0.145\sigma~M_{\odot}$ and $R_{\rm NS}=13.02\pm1.15\sigma~{\rm km}$, where $\sigma$ is the standard derivation. The constraints in $1\sigma$ and $2\sigma$ regions are adopted in this study. In our EOSs, the all EOSs except TOGASHI$+\pi$ can pass through the allowed $1\sigma$ regions. Considering the $2\sigma$ regions, TOGASHI$+\pi$ becomes a consistent model. This means that the current observation by {\it NICER} cannot reject all adopted EOSs, and we have waited for future observations by {\it NICER} for more constraints on EOS. 

\subsection{Constraint from the observations of photospheric radius expansion}
\label{subsec:d}
The observations of burst and quiescent phases in LMXBs enable us to probe the mass and radius. In particular, the observations of photospheric radius expansion (PRE) give their constraints due to their high brightness which is equal to the Eddington luminosity ($>10^{38}~{\rm erg~s^{-1}}$).  In this work, we adopt the analysis of six LMXBs of Ref.~\cite{2010ApJ...722...33S}, which results in a preferred radius of around 11--12 km. The TOGASHI is therefore good agreement with the allowed $1\sigma$ regions. Furthermore, TOGASHI$+\pi$ and TM1e+$\pi$ can pass through the allowed $2\sigma$ regions. We note that however the constraining regions of PRE observations might be changed due to some uncertain factors, such as the distance, compositions of the atmosphere, and the position of the photosphere~(for review, see Ref.~\cite{2016ARAA..54..401O}). 

\subsection{Constraint with the analysis from PSR J0437$-$4715}
\label{subsec:e}
The observed X-ray pulsations with XMM-Newton from the closest millisecond pulsar PSR J0437–4715 can probe the EOS~\cite{2013ApJ...762...96B}. The constraint in $3\sigma$ regions is adopted in this study. Moreover, we combine this constraint with its latest measured mass, which is measured in a radio timing method in the Parkes Pulsar Timing Array ($M_{\rm NS}=1.44\pm0.07~M_{\odot}$)~\cite{2016MNRAS.455.1751R}. Finally, the radius with this corresponding mass region is constrained as $R_{\rm NS}>10.9~{\rm km}$. All our models, including the smallest-radius model of the TOGASHI$+\pi$, match with the above constraints\footnote{Note that, however, the observational constraints of thermal pulses from PSR J0437$-$4715 in 2$\sigma$ regions exclude the TOGASHI and in 1$\sigma$ regions the TM1e as well, in regardless of the effect of pion condensation.}.

\subsection{Constraint from kHz QPO in LMXB 4U 0614$+$09}
\label{subsec:f}
In several LMXBs, quasi-periodic brightness oscillations (QPOs) have been observed. If the frequency is highly comparable to the orbital frequency of NS, the upper mass and radius could be determined at the same time. The highest frequency of the QPOs observed so far is 1.33 kHz in 4U 0614$+$09~\cite{2000ApJ...540.1049V}. Such a kHz QPO observation gives following loose constraints: $M_{\rm NS}<1.65~M_{\odot}$ and $R_{\rm NS}<14.7~{\rm km}$ without NS spin. Our EOSs are all consistent. If another kHz QPO with a higher frequency than that of 4U 0614+09 is observed in the future, the NS radius is preferred to be small with light stars.

\subsection{Possible constraint from thermal radiation of RX J1856.5$-$3754}
\label{subsec:g}
Seven isolated NSs are known to emit thermal X-ray radiation. In such a group called as {\it magnificent seven}, and the only object where the distance is measured is RX J1856.5$-$3754. Its blackbody radius $R_{\infty}$ is obtained, and therefore we could constrain on NS mass and radius~\cite{2002ApJ...564..981P}. We use the best fitting value $R_{\infty}=16.8~{\rm km}$ of the spectrum with blackbody emissions~\cite{2004NuPhS.132..560T}. The constraint of $R_{\rm NS}<R_{\infty}$ excludes the possibility of high-mass NSs in TM1, TM1$+\pi$, TM1e and TOGASHI. However, the constraint always allows the low-mass NSs to exist with all EOSs, although the distance is uncertain. Hence, compared with other constraints, it is hard to probe high-density EOS from kHz QPO observations.

\subsection{Possible constraint from the masses of PSR J0737$-$3039(B)}
\label{subsec:h}
PSR J0737$-$3039 is the only known double pulsar system, which may probe the EOS in high-density regions. The pulsar B, which is a lower-mass object in the system, is a light star with $M_{\rm NS}=1.249\pm0.001~M_{\odot}$. Such a light star is generally hard to be produced by a type-II supernova, which is triggered by a collapse of the Fe core of massive progenitor. As one of the scenarios to produce light NSs, an electron-capture supernova of an ONeMg core is suggested and has been confirmed from recent observation~\cite{2021NatAs.tmp..107H}. Setting the critical density of ONeMg core as $4.5\times10^9~{\rm g~cm^{-3}}$, the baryon mass of ONeMg core just before its collapse is estimated to be $1.375~M_{\odot}$. Then, the baryonic mass of formed NS is given as $M_{\rm B}\lesssim1.375~M_{\odot}$~\cite{2005MNRAS.361.1243P}. We show the gravitational mass corresponding to $M_{\rm B}=1.375~M_{\odot}$ for each EOS in Fig.~\ref{fig:mr}. We also show the measured mass of $M_{\rm NS}=1.249\pm0.001~M_{\odot}$, which should be higher than the gravitational mass with $M_{\rm B}=1.375~M_{\odot}$. We confirm that this condition is satisfied For all EOSs. Some previous studies consider the lower bounds of $M_{\rm B}$ derived from how the matter in the ONeMg core is ejected~(e.g., Ref.~\cite{2015PhLB..748..369M})\footnote{If there is no mass loss of the parent ONeMg core,  $M_{\rm B}\gtrsim1.366~M_{\odot}$~\cite{2005MNRAS.361.1243P}. By comparing it with the baryonic mass with $M_{\rm NS}=1.249\pm0.001~M_{\odot}$, we can test the EOS. As a result, all our EOSs are consistent with the constraints without mass loss. If we consider the mass loss, the estimated baryonic mass is decreased and the consistency is also changed. For example, one-dimensional core-collapse supernova simulation of ONeMg core shows the result of $M_{\rm B}=1.360\pm0.002~M_{\odot}$ with a mass lose
of $\sim0.015~M_{\odot}$~\cite{2006A&A...450..345K}. In that case, none of our adopted EOSs with $M_{\rm NS}=1.249\pm0.001~M_{\odot}$ passes through this allowed region.}, but this constraint has large uncertainties. Nevertheless, the baryonic-mass constraints from an electron-capture supernova can probe the NS EOS, although whether the pulsar B is formed in this scenario is still unknown.

\subsection{Constraints from Pulsar Glitches}
\label{subsec:i}
\begin{figure}[t]
    \centering
    \includegraphics[width=0.8\linewidth]{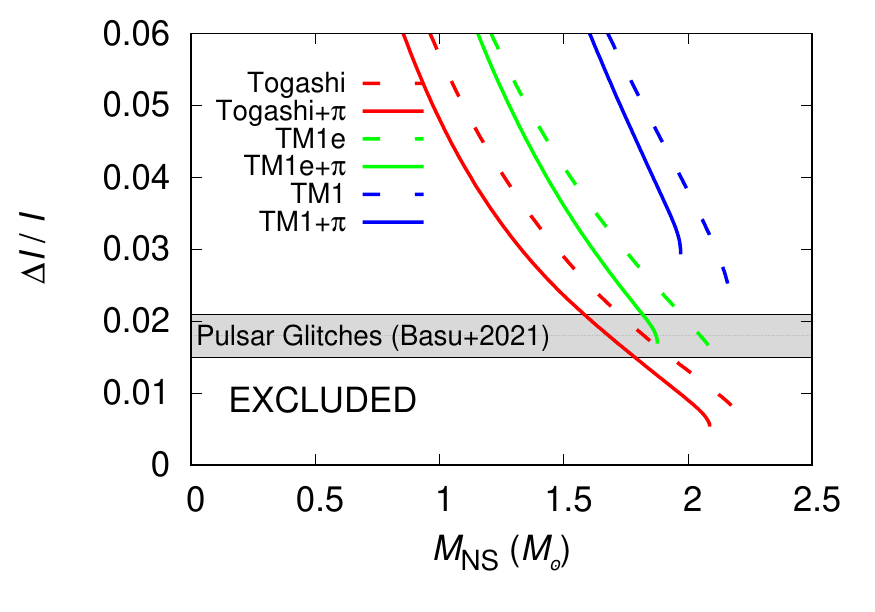}
    \caption{The fraction of crustal momentum of inertia in total one as a function of NS mass for each EOS. Grey band indicates the lower limits of recent pulsar-glitch observations~\cite{2021arXiv211106835B}.}
    \label{fig:pg}
\end{figure}
Recently, the analysis of 543 observed pulsar glitches, which are rapidly spin-up phenomena in rotating NSs, have been performed~\cite{2021arXiv211106835B}. They finally obtain the averaged rate where the pulsar's spin-down is reversed due to its glitch as $\dot{\nu}_g/|\dot{\nu}|=0.018\pm0.003$, where $\dot{\nu}_g$ is the glitch activity and $\dot{\nu}$ is the spin-down rate. Then , one can obtain the constraints on the fraction of the crustal moment of inertia $\Delta I$ in the total one $I$ as $\Delta I/I\gtrsim\dot{\nu}_g/|\dot{\nu}|$~\cite{1999PhRvL..83.3362L,2019PhRvC.100c5802L}. Important is that $\Delta I/I$ can be expressed as a function of $M_{\rm NS},R_{\rm NS},n_{\rm t},$ and $P_{\rm t}$, where $n_{\rm t}$ and $P_{\rm t}$ deonte the crust-core transition number density $n_{\rm t}$ and pressure $P_{\rm t}$, respectively. The empirical formula with high accuracy is given as follows~\cite{2000PhR...333..121L}:
\begin{eqnarray}
\frac{\Delta I}{I} = \frac{28\pi P_t R^3_{\rm NS}}{3M_{\rm NS}c^2}\frac{(1-1.67\beta-0.6\beta^2)}{\beta}\left[1+\frac{2P_t}{n_tm_bc^2}\frac{1+5\beta-14\beta^2}{\beta^2}\right]^{-1}~,~\label{eq:+}
\end{eqnarray}
where $\beta=GM_{\rm NS}/R_{\rm NS}c^2$ is the compactness parameter and $m_b$ is the averaged nucleon mass. In Fig.~\ref{fig:pg}, we show $\Delta I/I$ as a function of mass based on Eq.~\ref{eq:+}. As we see, the TOGASHI(+$\pi$) EOSs with high-mass regions are not preferred with the observation. On the other hand, other large-radius EOSs, except the TM1e+$\pi$ around maximum-mass regions, are consistent with the current glitch observations. We note that the pulsar-glitch constraint is much weaker than others because it can give only lower limits of $\Delta I/I$~\cite{1999PhRvL..83.3362L}, and always allow very low-mass NSs to exist as with sub-section~\ref{subsec:g}.

\section{Inputs for Cooling models}
\label{sec:nu}

\subsection{Cooling Processes}

As the slow cooling processes, we consider the neutrino emission of modified Urca, bremsstrahlung of nucleon-nucleon and electron-ion, electron-positron pair creation, photo-neutrino process, and plasmon decay processes~\cite{2001PhR...354....1Y}. In these processes, the modified Urca process and bremsstrahlung are dominant for the slow cooling scenario. These emissivities are approximately $10^{19-21}T_9^8~{\rm erg~cm^{-3}~s^{-1}}$, where $T_9$ is the local temperature in units of $10^9~{\rm K}$. For any slow cooling model, these processes are valid since they are always open.

In fast cooling processes, the nucleon DU process is considered with any EOS. The emissivity is given as approximately $10^{27}T_9^6~{\rm erg~cm^{-3}~s^{-1}}$, which is much higher than that of slow cooling processes. Once the DU process is open, it decreases the temperature in the core rapidly. However, the DU process is forbidden to occur if the momentum among reactant particles is not conserved in conventional NS matter. The threshold of proton fraction $Y_p^{e{\rm DU}}$ in the DU process via electrons is given as the following condition~\cite{1991PhRvL..66.2701L}:
\begin{eqnarray}
Y_p\ge Y_p^{e{\rm DU}} =
\begin{cases}
 1/9 & \text{if}~~Y_{\mu} = 0 \\
 0.1477 & \text{if}~~Y_{\mu} = Y_e\\
 \left[1+\left(1 + x_e^{1/3}\right)^3\right]^{-1} & \text{otherwise} ~~,
\end{cases}
\label{eq:eq3}
\end{eqnarray}
where $Y_e$ is electron fraction, $Y_{\mu}$ is muon fraction, and $x_e = Y_e/\left(Y_e + Y_{\mu}\right)$. 

\begin{figure}[t]
    \centering
    \includegraphics[width=0.8\linewidth]{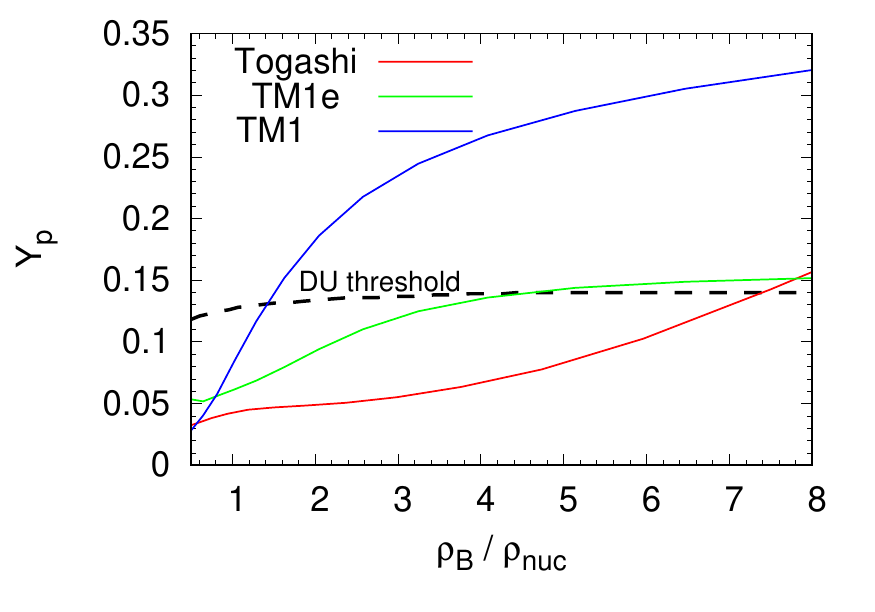}
    \caption{$Y_p$ distribution as a function of the density with adopted EOSs with standard-nuclear matter. For the threshold of the DU process, we fix as the TM1e for calculating $x_e$ in Eq.~(\ref{eq:eq3}).}
    \label{fig:yp}
\end{figure}

To see the onset density of the DU process, we show the density dependence of $Y_p$ in Fig.~\ref{fig:yp}. The high-symmetry-energy EOS, such as the TM1, has high $Y_p$ values even with relatively low-density regions. This implies that the DU process occurs even with low mass stars. In low-symmetry energy EOS, $Y_p$ is low even with high-density regions. That is, the DU process occurs with only high mass stars or does not occur. In the case of the TOGASHI, the DU process is forbidden in NSs. That is why another fast cooling process is required for cold NS observations~\cite{2019PTEP.2019k3E01D}. In adopted EOSs without pion condensation, the threshold mass is given as follows: $M_{\rm DU}=2.06 M_{\odot}$ for the TM1e, $M_{\rm DU}=0.77 M_{\odot}$ for the TM1, and $M_{\rm DU}>M_{\rm max}=2.21 M_{\odot}$ for the TOGASHI. Hence, the higher-$M_{\rm DU}$ value certainly corresponds to the lower symmetry energy, simply $L$. $M_{\rm DU}$ could be changed because of the softening EOS due to pion condensation, but the effect is negligible in our models.

If the NS matter includes an exotic state beyond $npe\mu$, another fast cooling process may occur. In this work, we adopt the pion Urca process modeled by Ref.~\cite{1993PThPS.112..221M}. They suggest that an extra cooling mechanism of quasi($\eta$)-particle Urca process could occur in pion condensation phases. Considering the $\eta$-particles and leptons in thermal equilibrium, the pion Urca process can be expressed as follows:
\begin{eqnarray}
\eta(\bm{p}) &\rightarrow& \eta(\bm{p^{\prime}}) + l + \bar{\nu}_l  \nonumber,\\
\eta(\bm{p})+ l &\rightarrow& \eta(\bm{p^{\prime}}) + \nu_l~~, \label{eq:eq4}
\end{eqnarray}
where $\bm{p}$ and $\bm{p^{\prime}}$ denote momentum of $\eta$ particles. According to Ref.~\cite{1993PThPS.112..221M,1994ApJ...431..309U}, the neutrino emissivity is  around $10^{24-25}T_9^6~{\rm erg~cm^{-3}~s^{-1}}$ (see also Fig.~2 in Ref.~\cite{2018IJMPE..2750067M}). Although the concrete coefficient of the emissivity is different between $\pi^c$ and $\pi^0$-$\pi^c$ phase, this pion Urca process is clearly stronger than slow cooling processes.

\begin{figure}[t]
    \centering
    \includegraphics[width=0.8\linewidth]{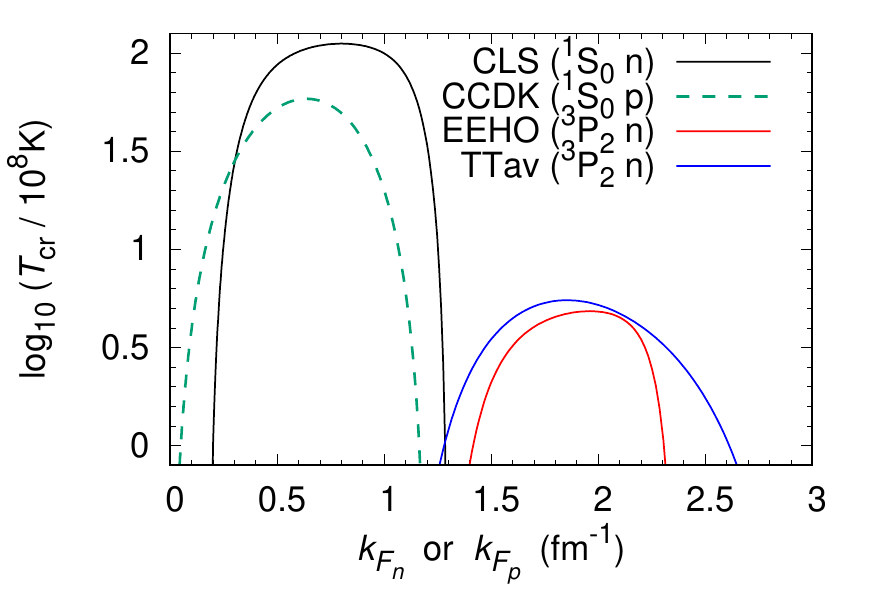}
    \caption{Density Dependence of superfluid transition temperature adopted in this study. The horizontal axis indicates {\em Fermi}-wave number of neutrons with solid curves, while of protons with dashed curves.}
    \label{fig:super}
\end{figure}

Nucleon superfluidity is also important for NS cooling curves since the temperature ($\ll1~{\rm MeV}$) may become lower than the superfluid transition temperature $T_{\rm cr}$. As the effect of superfluidity on cooling curves, we consider two physical processes (for review, see Ref.~\cite{2013arXiv1302.6626P}): One is the suppression of neutrino emission,  specific heat, and thermal conductivity. The other one is the PBF processes. The efficiency of these effects depends on the density dependence of $T_{\rm cr}$, but there are still unknown due to the uncertainties of nuclear interactions. In a lot of nucleon superfluid models~(e.g., Ref.~\cite{2015PhRvC..91a5806H}), we choose following superfluid models: CLS~\cite{2006PhRvC..74f4301C,2009PhRvC..80d5802G} and CCDK~\cite{1993NuPhA.555...59C} for neutrons and protons ${}^1S_0$ channels, respectively. For the neutron superfluidity in the ${}^3P_2$ channel, the DU process and the pion Urca process are greatly suppressed~(e.g., Ref.~\cite{1998PhR...292....1T}). So, we choose weak and strong pairing models: EEHO~\cite{1996NuPhA.607..425E} and TTav~\cite{2004PThPh.112...37T}, respectively. We show the density dependence of superfluid transition temperature $T_{\rm cr}$ in Fig.~\ref{fig:super}. Compared with EEHO, TTav has a higher superfluid effect in higher-density regions. Hence, pairing effects on cooling curves with high-mass NSs are higher with TTav than EEHO. 

\subsection{Basic Formulation}

In this work, we use the spherically symmetric relativistic stellar evolutionary code~\cite{1984ApJ...278..813F}. When there is no heating process or convection, the basic equations to describe the temperature and luminosity of isolated NSs are as follows ($c=G=1$)~\cite{1977ApJ...212..825T,2018A&A...609A..74P}:
\begin{eqnarray}
\frac{\partial (L_{r}e^{2\phi})}{\partial M_{r}} & = &
      -e^{2\phi}\left(\varepsilon_{\nu} + e^{-\phi}C_V\frac{\partial T}{\partial t}
      \right)~, \label{eq:eq5} \\
      \frac{\partial \ln T}{\partial \ln P} & = & \frac{3}{16\pi} \frac{\kappa L_r P}{M_{{\rm t}r}aT^4}\frac{\rho_0}{\rho}\left( 1 + \frac{P}{\rho} \right)^{-1}\times \nonumber \\
&&\left( 1 + \frac{4 \pi r^3 P}{M_{{\rm t}r}} \right)^{-1}\left(1-\frac{2M_{{\rm t}r}}{r}\right)^{1/2} \nonumber \\
&& + \left[1 - \left( 1 + \frac{P}{\rho} \right)^{-1} \right]~, \label{eq:eq6}
\end{eqnarray}
where $M_{tr}$ and $M_r$ are gravitational and rest masses enclosed in a radius $r$; $\rho$ and $\rho_0$ denote the total mass-energy and rest mass densities; $P$, $T$, and $L_r$ are the pressure, local temperature, and local photon luminosity, respectively, $\varepsilon_\nu$ denotes the energy loss rate by neutrino emission; $\phi$ is the gravitational potential in unit mass; $a$ is the Stefan-Boltzmann constant; $C_V$ is the specific heat; $\kappa$ is the opacity. As the boundary condition, we impose the radiative zero boundary condition at sufficiently closed area to the photosphere~\cite{1984ApJ...278..813F}. By solving Eqs.~(\ref{eq:eq5}), (\ref{eq:eq6}) and the TOV equations by Henyey method, we can obtain the time evolution of the luminosity, and the surface temperature via Stefan--Boltzmann law.

NS surface composition is one of the important factors for describing cooling curves~(for review, see Ref.~\cite{2021PhR...919....1B}). If there are more light elements onto the NS surface, the surface temperature is generally higher at the neutrino cooling stage. Meanwhile, this trend becomes the opposite at the photon cooling stage. In this work, we consider two extreme cases: pure Ni surface and pure He surface with $M_{\rm env}/M_{\rm NS} = 10^{-7}$, where $M_{\rm env}$ is the envelope mass and physically up to $\sim10^{-7}$ times of NS gravitational mass $M_{\rm NS}$~\cite{1997A&A...323..415P}. For the opacity models, we consider the radiative opacity~\cite{1999ApJ...524.1014S} and conductive opacity composed of mainly electrons~\cite{2015SSRv..191..239P} and neutrons~\cite{2001A&A...374..151B}. 

As temperature observations of isolated NSs, we adopt the 18 data points in Ref.~\cite{2017IJMPE..2650015L}. The data include the observations of PSR J0205+6449 in supernova remnant 3C58 and RX J0007.0+7302 in CTA 1, which are beyond minimal cooling scenario~\cite{2004ApJS..155..623P}. Hence, these observations are strong evidence for fast cooling processes. Currently, the only uppers limits of the surface temperature are known for them, so the temperature observations still include large uncertainties. Nevertheless, X-ray observations of NS temperature have been recently proceeded rapidly as a representative of {\it NICER}~(e.g., Ref.~\cite{2021arXiv210506980R} for PSR J0740+6620). Thus, if the X-ray observations make progress in the future, we might specify which kind of fast cooling process occurs with the use of the accurately measured temperature data.

\begin{figure}[t]
 \centering\hspace*{-0.5cm}
    \includegraphics[width=\linewidth]{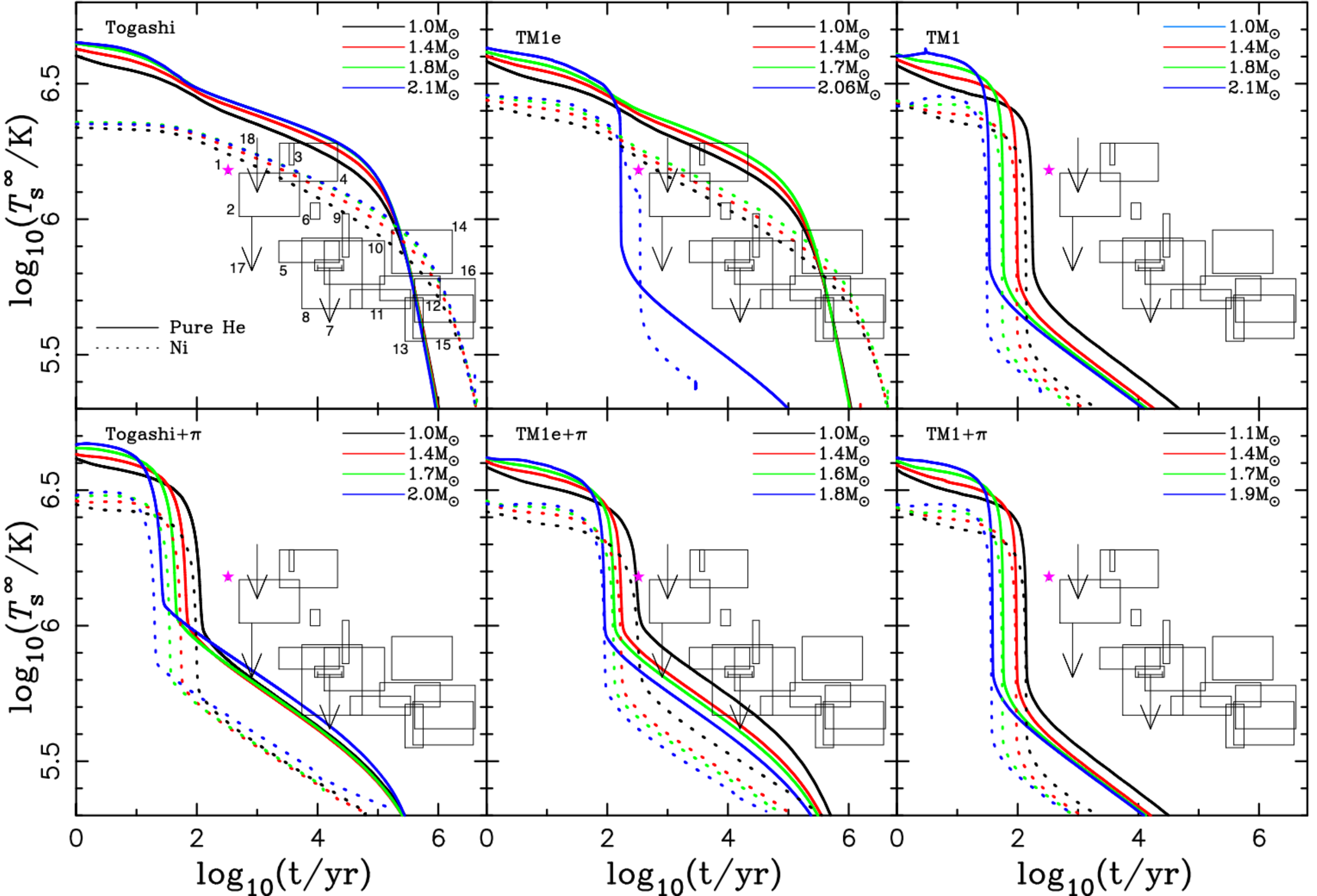}
\caption{Time evolution of redshifted-surface temperature of isolated NSs, cooling curves, without nucleon superfluidity. Upper panel: pion condensation is not included. The adopted EOSs are as follows: TOGASHI(left), TM1e(middle), and TM1(right). The pion condensation is considered in lower panel. Solid curves indicates with the He envelope ($M_{\rm env}/M_{\rm NS}=10^{-7}$) while dotted curves with the Ni envelope. The chosen mass is different from colors of cooling curves. The data of cooling observations (1--18 in left-top panel) are taken from Ref.~\cite{2017IJMPE..2650015L}.}
    \label{fig:nosf}
    \end{figure}
     \begin{figure}[t]
    \centering\hspace*{-1.5cm}
    \includegraphics[width=0.9\linewidth,angle=-90]{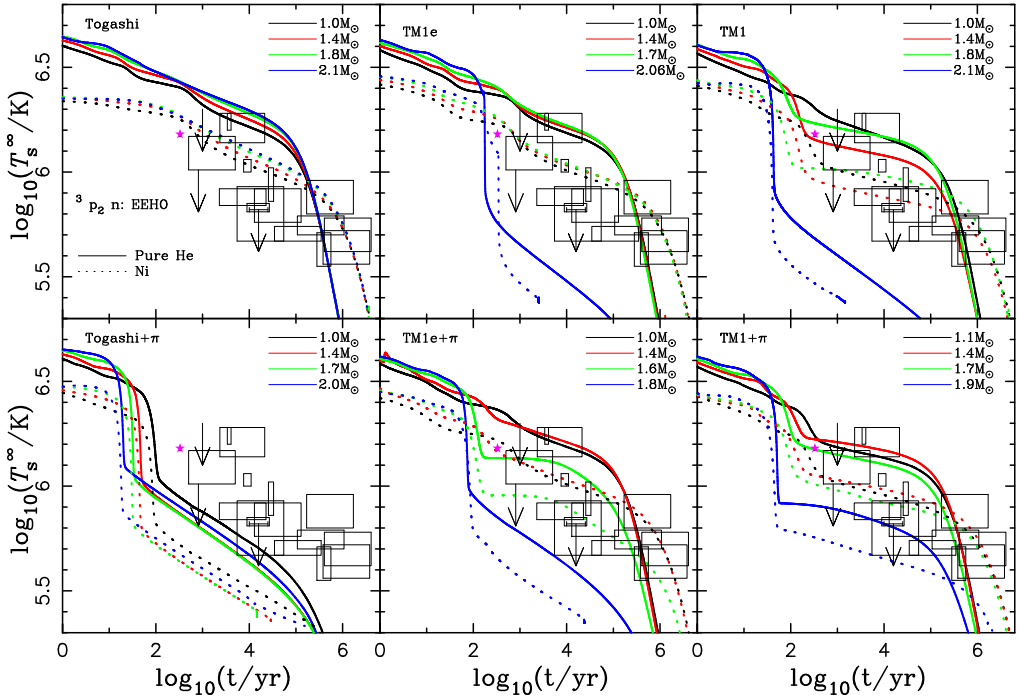}
    \caption{Same as Fig.~\ref{fig:nosf}, but considering the effect of nucleon superfluidity: CLS, CCDK, and EEHO for neutrons ${}^1S_0$, protons ${}^1S_0$, and neutrons ${}^3P_2$ channels, respectively.}
    \label{fig:eeho}
\end{figure}
 \begin{figure}[t]
    \centering\hspace*{-1.5cm}
    \includegraphics[width=0.9\linewidth,angle=-90]{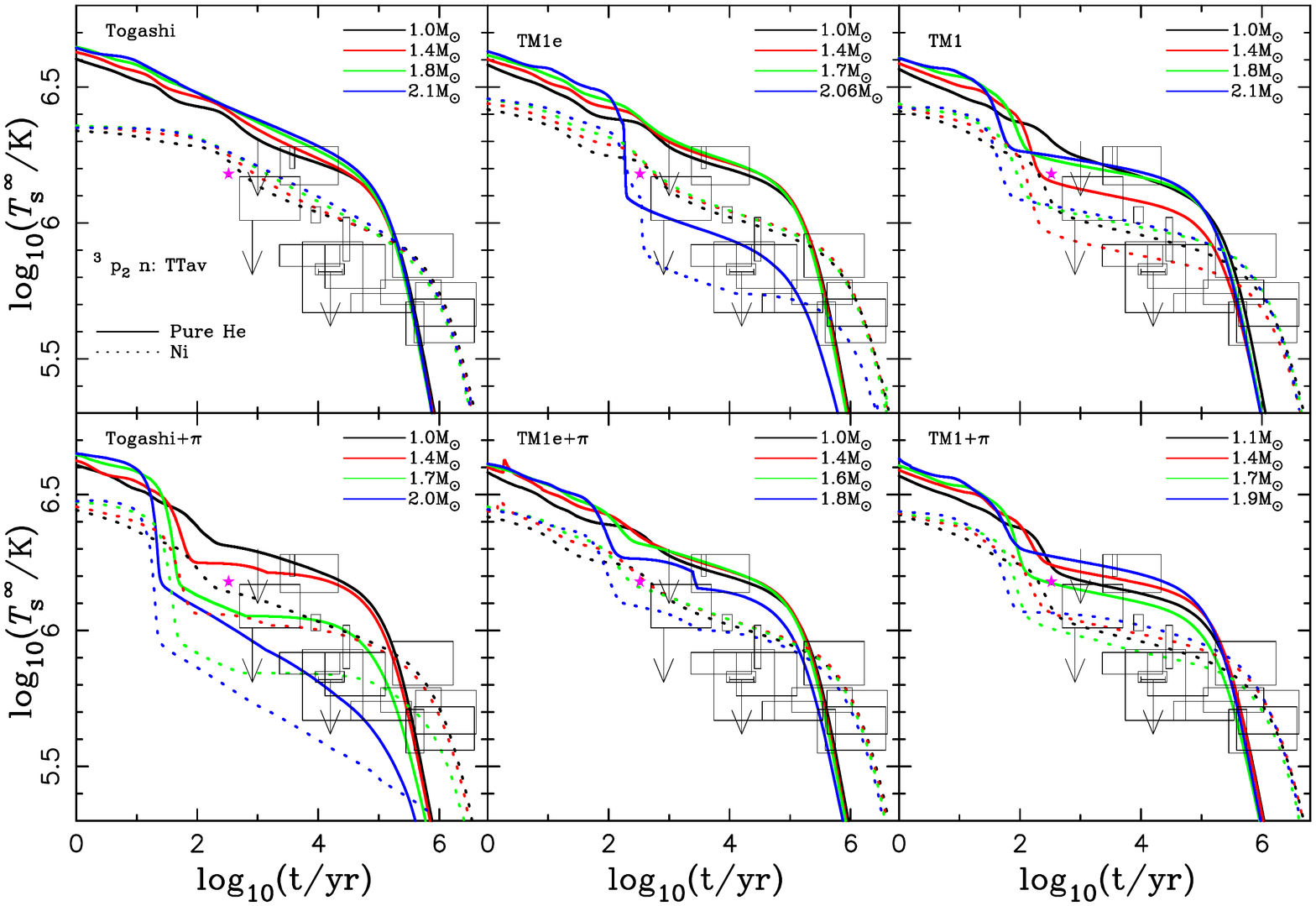}
    \caption{Same as Fig.~\ref{fig:eeho}, but with the TTav for neutrons ${}^3P_2$ channel.}
    \label{fig:ttav}
\end{figure}

\section{Results}
\label{sec:result}

First of all, we show the simple cooling curves which do not include the pion condensation and nucleon superfluid effects in the upper panel of Fig.~\ref{fig:nosf}. The cooling curves with the TOGASHI do not show fast cooling even with $2.1 M_{\odot}$ stars because of low symmetry energy enough to prohibit the DU process. Isolated NSs with the TM1 cause the fast cooling with $M\ge1.0M_{\odot}$ because of the high symmetry energy. These two extreme cases cannot explain most cooling observations~(see also Ref.~\cite{2019PTEP.2019k3E01D}). For the TM1e, the DU process occurs with $M=2.06 M_{\odot}$, but it does not occur with $M=1.7 M_{\odot}$. The cooling observations between two kinds of cooling curves seem to be on the masses of (1.7--2.06)$M_{\odot}$, but considering that the DU process is very sensitive to the mass~(e.g., see Fig.~9 in Ref.~\cite{2017IJMPE..2650015L}), reproducing such observations without nucleon superfluidity seems to be difficult, with both heavy and light envelope models. 

Considering the pion condensation, cooling curves of all models, which are shown in bottom panel of Fig.~\ref{fig:nosf}, become inconsistent with high-temperature observations due to the strong pion Urca process. Even with the low-mass stars, the pion Urca process is valid since the pion condensation appears with the relatively low density $\rho_{\rm B}=1.6\rho_{\rm nuc}$, where the mass is very small ($\lesssim 1 M_{\odot}$). For the TOGASHI$+\pi$, the fast cooling process can lower the surface temperature, whose behavior cannot be seen in the original TOGASHI, but the pion Urca process is too strong to explain most stars with high-temperature regions.

Next, we show the cooling curves with nucleon superfluidity in Fig.~\ref{fig:eeho} and Fig.~\ref{fig:ttav}. In both the DU process and pion Urca process, the parameter of special importance is the strength of the neutron superfluid model in the ${}^3P_2$ channel, which greatly contributes to the cooling suppression in theory~\cite{1972PThPh..48.1517T}. In all cooling curves with fast cooling processes, if the neutron superfluid model in the ${}^3P_2$ channel is stronger, the cooling curves move to higher-temperature regions due to higher cooling suppression. In this work, since we focus on the possibility of pion condensation in NSs, we discuss the cooling curves mainly with the pion condensation.

For the TM1, the DU process occurs with $M_{\rm NS}\ge1.0 M_{\odot}$ as we can see the cooling curves in the top-right panel of Fig.~\ref{fig:nosf}. For the TM1$+\pi$,
the pion Urca process occurs, but it seems to be hidden by the stronger DU process as we see bottom-right panel of Fig.~\ref{fig:nosf}. Hence, the additional fast cooling process in exotic matter is required for the high-symmetry-energy EOS. Rather, the pion condensation is not preferred for the high-symmetry-energy EOS because such a model is sensitive to the maximum mass due to the softening effect.

For the TM1e, the fast cooling process derived from the DU process occurs with $M_{\rm NS}\ge2.06 M_{\odot}$. For the TM1e$+\pi$, the pion Urca process also occurs, but it could be hidden by the stronger DU process with $M_{\rm NS}\ge2.06 M_{\odot}$. However, with the masses $M_{\rm NS}<2.06 M_{\odot}$, the pion Urca process is dominant for cooling curves. Considering the nucleon superfluidity, the pion Urca process becomes milder. The EEHO model of neutron superfluidity in the ${}^3P_2$ channel can be well fitted with cooling observations, but the TTav cannot explain cold NSs because the cooling suppression is too strong. Hence, mild superfluid models in the ${}^3P_2$ channel seem to be better for cooling observations.

For the TOGASHI$+\pi$, the pion Urca process is dominant for cooling curves with any masses. With the EEHO for neutrons ${}^3P_2$ superfluidity, the cooling suppression is too weak to explain some warm stars. But with the TTav, the cooling observations can be reproduced. Hence, by considering the strong neutrons ${}^3P_2$ superfluidity within the high-density regions, such a model with low symmetry energy is consistent with cooling observations. Since the low-symmetry-energy EOS is not softened so much in high-density regions $\rho_{\rm B}\gg\rho_{\rm nuc}$, such a cooling model with pion condensation could be one of the candidates to solve the problem of $2 M_{\odot}$ and cold cooling observations.

As above, if the EOS is different, the efficiency of cooling suppression by nucleon superfluidity is also different. Our results show that the standard-nuclear EOSs with lower symmetry energy need stronger neutron ${}^3P_2$ superfluidity for cooling observations, not only to reproduce $2 M_{\odot}$ observations as shown in Fig.~\ref{fig:mr}. Therefore, the low-symmetry-energy EOS which is enough to prohibit the DU process could be modified for solving both problems of $2 M_{\odot}$ and cold cooling observations simultaneously, by considering the pion condensation and strong neutron superfluidity in the ${}^3P_2$ channel. 

\section{Conclusion}
\label{sec:conc}

We studied thermal evolution of isolated NS with constructed EOSs with pion condensation, focusing on the softness of standard-nuclear EOS. As a result, the TOGASHI$+\pi$ is in good agreement with the $2 M_{\odot}$ observations and cold cooling observations. The former is based on the softness of EOS with the standard-nuclear matter, which can be associated with the symmetry energy. In high-density regions with $\rho_{\rm B}\gg \rho_{\rm nuc}$, the EOS with lower symmetry energy does not become softer so much by the pion condensation, and this enables such an EOS to support $2 M_{\odot}$. The latter is connected with the occurrence of fast cooling processes and the neutron superfluidity in the ${}^3P_2$ channel. The low-symmetry-energy EOS which is enough to prohibit the DU process, such as the TOGASHI, requires another fast cooling process. Then, we considered the pion Urca process as one of the candidates for them. As a result, most cooling observations could be reproduced with the strong neutrons superfluidity in the ${}^3P_2$ channel. As one of such consistent cooling models, TOGASHI$+\pi$, associated with the TTav neutron ${}^3P_2$ superfluid model, was present in this paper.

According to the recent experiment of the Gamow-Teller Giant Resonance in neutron-rich double magic nucleus ${}^{132}{\rm Sn}$ resulted in the following constraint of $\tilde{g}^{\prime}_{NN}=0.68\pm0.07$~\cite{2018PhRvL.121m2501Y}. The value of $\tilde{g}^{\prime}_{NN}$ is larger than our value of $\tilde{g}^{\prime}=0.5$. However, the universality is shown to be against another experiment of the quenching on the Gamow-Teller transitions~\cite{1999PhLB..455...25S}. Making the universality milder, we finally obtain $\tilde{g}^{\prime}=0.5$--$0.6$~\cite{1993PThPS.112..159M}, which is lower than $\tilde{g}^{\prime}_{NN}\simeq0.68$. Thus, our choice of $\tilde{g}^{\prime}=0.5$ would be justified from the nuclear experiment, although it still remains some uncertainties. 

The uncertainties of $\tilde{g}^{\prime}$ affect cooling curves; if $\tilde{g}^{\prime}$ is greater, the pion condensation occurs with higher-density regions. The threshold mass of the pion Urca process becomes higher and therefore only heavier NS cools rapidly (see Ref.~\cite{1994ApJ...424..846R} for the comparison between $\tilde{g}^{\prime}=0.5$ and $0.6$). Hence, it is worth making other EOSs with different $\tilde{g}^{\prime}$ and checking them with temperature observations. Nevertheless,  unless the pion condensation is prohibited with any mass due to being larger $\tilde{g}^{\prime}$, the cooling scenario of low-symmetry-energy EOSs would not be changed, considering the suppression of neutrino emissivities by strong neutron superfluidity. 

We note that the pion condensation scenario is not a unique one to solve the problem of low-symmetry-energy EOSs. For example, hyperon mixing is also another candidate because hyperon DU process occurs and may cool NSs rapidly. In particular, $\Lambda$-hyperon DU processes (e.g., $\Lambda\rightarrow p+e+\bar{\nu}_e$) really works because of weak $\Lambda\Lambda$ pairing gap~\cite{2001PhRvL..87u2502T,2006PThPh.115..355T}. For another example, the kaon condensation could be another good candidate because kaon-Urca process occurs and may cool NSs rapidly as well. Recently, such exotic-matter EOSs with strangeness have been well constructed and can support $2M_{\odot}$ stars due to three-body force (e.g., Ref.~\cite{2016PhRvC..93c5808T} for hyperon-mixed matter, and Ref.~\cite{2021arXiv210603449M} for kaon-condensation matter). We are going to investigate cooling behavior in such EOSs with strangeness. We hope that further cooling observations are beneficial to explore the NS matter.

\small
\section*{Acknowledgments}
We thank Prof. Takumi Muto for careful reading of our manuscript and giving several useful comments on pion condensation. A.D. is partially supported by RIKEN iTHEMS Program. The work of H.L. has been supported financially by the National Natural Science Foundation of China under No. 11803026 and Xinjiang Natural Science Foundation under No. 2020D01C063.
T.N. wishes to acknowledge the support from the Discretionary Budget of the
President of Kurume Institute of Technology.

\bibliography{ref}
\bibliographystyle{ws-ijmpe}

\end{document}